\documentstyle[epsf,eqsecnum,floats,preprint,aps]{revtex}

\tighten

\input epsf
\begin{document}

\newcommand{\be}{\begin{equation}}
\newcommand{\ee}{\end{equation}}
\newcommand{\bea}{\begin{eqnarray}}
\newcommand{\eea}{\end{eqnarray}}
\newcommand{\PSbox}[3]{\mbox{\rule{0in}{#3}\includegraphics{#1}\hspace{#2}}}

\def\5M{M^3_{(5)}}
\def\4M{M^2_{(4)}}

\overfullrule=0pt
\def\Int{\int_{r_H}^\infty}
\def\d{\partial}
\def\e{\epsilon}
\def\M{{\cal M}}
\def\high{\vphantom{\Biggl(}\displaystyle}
\catcode`@=11
\def\@versim#1#2{\lower.7\p@\vbox{\baselineskip\z@skip\lineskip-.5\p@
    \ialign{$\m@th#1\hfil##\hfil$\crcr#2\crcr\sim\crcr}}}
\def\simge{\mathrel{\mathpalette\@versim>}} %
\def\simle{\mathrel{\mathpalette\@versim<}} %
\def\sun{\hbox{$\odot$}}
\catcode`@=12 

\rightline{astro-ph/0307034}
\vskip 2cm

\setcounter{footnote}{0}

\begin{center}
\large{\bf Differentiating between Modified Gravity and Dark Energy}
\ \\
\ \\
\normalsize{Arthur Lue$^1$\footnote{E-mail: lue@bifur.cwru.edu},
  Rom\'an Scoccimarro$^2$\footnote{E-mail: rs123@nyu.edu},
  and Glenn Starkman$^1$\footnote{E-mail: starkman@balin.cwru.edu}}
\ \\
\ \\
${}^{1}$\small{\em Center for Education and Research in Cosmology
  and Astrophysics\\
Department of Physics\\
Case Western Reserve University\\
Cleveland, OH 44106--7079}
\ \\
\ \\
${}^{2}$\small{\em Center for Cosmology and Particle Physics\\
Department of Physics\\
New York University \\
New York, NY 10003}

\end{center}

\begin{abstract}

\noindent
The nature of the fuel that drives today's cosmic acceleration is an
open and tantalizing mystery.  We entertain the suggestion that the
acceleration is not the manifestation of yet another new ingredient in
the cosmic gas tank, but rather a signal of our first real lack of
understanding of gravitational physics.  By requiring that the
underlying gravity theory respects Birkhoff's law, we derive the
modified gravitational force-law necessary to generate any given
cosmology, without reference to the fundamental theory, revealing
modifications of gravity at scales typically much smaller than today's
horizon. We discuss how through these modifications, the growth of
density perturbations, the late-time integrated Sachs--Wolfe effect,
 and even solar-system measurements may be sensitive to whether
today's cosmic acceleration is generated by dark energy or modified
gravitational dynamics, and are subject to imminent observational
discrimination.  We argue how these conclusions can be more
generic, and probably not dependent on the validity of Birkhoff's law.
\end{abstract}

\setcounter{page}{0}
\thispagestyle{empty}
\maketitle

\eject

\vfill

\baselineskip 18pt plus 2pt minus 2pt

\section{Introduction}

The discovery of a contemporary cosmic acceleration
\cite{Perlmutter:1998np,Riess:1998cb} is one of the most profound
scientific observations of the latter part of the last century.  What
drives that acceleration remains an open and tantalizing question.  A
``conventional'' explanation exists for the cause of that
acceleration: vacuum energy provides the necessary repulsive gravity
in general relativity to drive an accelerated expansion of the
universe.  Variations on this vacuum-energy theme, e.g. quintessence,
promote the energy density to the potential energy density of a
dynamical field.  Such additions to the roster of cosmic sources of
energy-momentum are collectively referred to as dark energy.
Accounted for as such, dark energy would constitute the majority of
the energy density of the universe today.

However, what if one were to view the current cosmic expansion not as
yet another ingredient in our already complex cosmic soup, but rather
as a signal of our first real lack of understanding of gravitational
physics?  (Although others have argued that dark matter was that first
signal of post-Einsteinian gravity \cite{MOND,Sanders:2002pf}, we will
not address ourselves here to that possibility.)  An instructive
example along that direction is the braneworld theory
\cite{Dvali:2000hr} of Dvali, Gabadadze, and Porrati (DGP).  In this
theory, gravity appears four-dimensional at short distances but is
altered at large distances through the slow evaporation of the
graviton off our four-dimensional braneworld universe into an unseen,
yet large, fifth dimension
\cite{Dvali:2000hr,Dvali:2001gm,Dvali:2001gx}.  DGP gravity provides
an alternative explanation for today's cosmic acceleration
\cite{Deffayet,Deffayet:2001pu}: just as gravity is conventional
four-dimensional gravity at short scales and appears five-dimensional
at large distance scales, so too the Hubble scale, $H(t)$, evolves by
the conventional Friedmann equation at high Hubble scales but
saturates at a fixed value as $H(t)$ approaches a value equivalent to
the inverse of the crossover distance between four and
five-dimensional behavior, $r_c^{-1}$.  Thus, if one were to set that
crossover distance scale to be on the order of $H_0^{-1}$, where $H_0$
is today's Hubble scale, DGP could account for today's cosmic
acceleration in terms of the existence of extra dimensions and a
modifications of the laws of gravity.

We would naively expect not to be able to probe this extra dimension
at distances much smaller than the crossover scale $r_c = H_0^{-1}$.
However, in DGP, although gravity is four-dimensional at distances
shorter than $r_c$, it is not four-dimensional Einstein gravity -- it
is augmented by the presence of an ultra-light gravitational scalar.
One only recovers Einstein gravity in a subtle fashion
\cite{Deffayet:2001uk,Lue:2001gc,Gruzinov:2001hp,Porrati:2002cp}, and
a marked departure from Einstein gravity persists down to distances
much shorter than $r_c$.  For example, for $r_c=H_0^{-1}$ and a
central mass source of Schwarzschild radius $r_g$, significant and
cosmologically-sensitive deviations from Einstein gravity occur at
distances greater than
\cite{Gruzinov:2001hp,Porrati:2002cp,Lue:2002sw,Dvali:2002vf}
\be
     r_* = \left(r_g r_c^2 \right)^{1/3}  =
     \left({r_g\over H_0^2}\right)^{1/3}\ .
\ee
Thus a marked departure from conventional physics persists down to
scales much smaller than the distance at which the extra dimension is
naively hidden, or for our discussion here, the distance at which the
Friedmann equation was modified to account for accelerated cosmic
expansion. (Other theories have since shown how alternative modifications
of gravity at large distances can lead to late-time acceleration without
dark energy \cite{Damour:2002wu,Carroll:2003wy}.)

Recently, appeals have been made to a direct empirical modification of
the Friedmann equation in order to explain cosmic acceleration without
dark energy \cite{Freese:2002sq,Freese:2002gv,Dvali:2003rk},
broadening the theme explicitly and completely realized by DGP
braneworlds.  Unfortunately, fully self-consistent models implementing
these ideas have yet to
be found that can reproduce the desired general modifications to the Friedmann
equation, making it difficult to establish the full consequence of the
proposed new physics.  An important question to be asked is whether
gravity theories that produce such modified Friedmann equations can be
devoid of observable consequences other than late-time acceleration?
There may be something to be learned from the example of DGP gravity.
Will these new theories generically lead to similar deviations from
Einstein gravity at scales much smaller than today's Hubble radius?

Pursuing this line of thought without an underlying model is
difficult, but not altogether impossible.  In this paper, we allow
ourselves an assumption about the structure of a possible modified
theory of gravity, taking the constraint of Birkhoff's law as the
example, and show how with that assumption alone, one may
kinematically ascertain gravitational force interactions, {\em
avoiding} reference to fundamental dynamics.  We then describe how
modification of cosmology at today's Hubble scale can naturally affect
gravitational physics at much smaller (e.g.  astrophysical, and even
under certain circumstances solar system) distance scales.  It is
worth noting that DGP gravity does not respect Birkhoff's law, and we
conclude with a discussion on how short-distance modifications of
gravitational interactions may be a general consequence of gravity
modified at cosmological scales.

\section{The Friedmann Equation and the Force Law}

The premise of our approach is that the static, Schwarzschild-like
metric, or more specifically its geodesics, may completely determine
cosmological evolution, i.e., that the cosmological evolution is
driven not by some dark energy background, but by the matter content
itself.  This should be a familiar premise -- it underlies the usual
undergraduate-level derivation of the Friedmann equation from Newton's
laws of motion and his Law of Gravitation.  One posits that in a
homogeneous isotropic dust-filled universe, the acceleration of a
comoving (geodesic) observer a distance $r$ from some arbitrary
origin, is determined by the gravitational pull of the matter interior
to the sphere of radius $r$.  One then derives an equation for the
evolution of $r$ with time. That one arrives at the precisely correct
General-Relativistic Friedmann equation, may seem surprisingly
coincidental but is really a consequence of homogeneity, energy
conservation and Birkhoff's
law.  Thus in the standard cosmology, Newton's Law of Gravitation, or
now its General Relativistic generalization, and the observed scale
factor evolution are used to extract the stress-energy content as a
function of time.  The approach here will be to fix the form of the
stress energy to be that of dust (non-relativistic matter) and use the
observed evolution of the scale factor to extract a new generalization
of Newton's Law.

We take the cosmological metric to be the Robertson--Walker metric
(flat space, for convenience)
\bea
     ds^2 &=& dt^2 - a^2(t)~\delta_{ij}dx^idx^j   \nonumber \\
     &=& dt^2 - a^2(t)\left[d\lambda^2 + \lambda^2d\Omega\right]\ .
\label{cosmo}
\eea
We imagine then that we are given a complete cosmological evolution,
i.e.,  a specified cosmological scale factor evolution, $a(t)$.
(The understanding being that $a(t)$ is determined on the past
light cone by observations, and that the assumption of spatial
homogeneity carries this to the whole space.)  If we were to solve the
Einstein equations for this metric, we would be forced to specify a
matter content and an equation of state as a function of time.  Such
an equation of state would not normally be matter dominated, and we
would be forced to consider the non-dust component to be a sort of
dark energy.  However, we will instead {\em presume} that the scale
factor $a(t)$ {\em is} the result of a pure dust configuration, and
ask what modifications of gravity would be necessary to yield the
given $a(t)$.

Such a prescription for modifying the Einstein equations is not
unique.  However, we can require that this hypothetical new gravity
theory obeys a generalization of Birkhoff's law: for any test
particle outside a spherically symmetric matter source, the metric
observed by that test particle is equivalent to that of a point source
of the same mass located at the center of the sphere.  With that one
restriction on the theory, we can deduce completely the
Schwarzschild-like metric of the new hypothetical gravity theory that
gives us the prescribed cosmological evolution with a dust-filled
universe.

The procedure for determining the metric from the cosmological
evolution is as follows.  Consider a uniform sphere of dust.  Imagine
that the evolution inside the sphere is exactly cosmological, while
outside the sphere is empty space, whose metric (given Birkhoff's law)
is Schwarzschild-like (as defined by the metric Eq.~(\ref{line-static})
below).  The mass of the matter source (as determined by the form of
the metric at short distances) is unchanged throughout its
time-evolution. The surface of the spherical mass therefore charts out
the metric through all of space as the sphere expands with time, so
long as we demand that the cosmological metric just inside the surface
of the sphere smoothly matches the Schwarzschild solution just
outside.  In order to see how the metric depends on the mass of the
central source, we just take a sphere of dust of a different initial
size, and watch its surface chart out a new metric.

We start with a Schwarzschild-like metric in the usual form
\be
     ds^2 = g_{00}(r)dT^2 - g_{rr}(r)dr^2 - r^2d\Omega\ ,
\label{line-static}
\ee
and rewrite it in the new form
\be
     ds^2 = N^2(t,\lambda)dt^2 - a^2(t)d\lambda^2 - r^2d\Omega\ ,
\label{line-cosmo}
\ee
where now $r = r(t,\lambda)$ and $a(t)$ is the given cosmological
evolution.  In order to determine the coordinate
transformation $T=f(t,\lambda)$ and $r(t,\lambda)$, we equate the forms
Eq.~(\ref{line-static}) and Eq.~(\ref{line-cosmo}):
\bea
     N^2 &=& g_{00}\dot{f}^2 - g_{rr}\dot{r}^2
\label{trans1}     \\
     0 &=& g_{00}\dot{f}f' - g_{rr}\dot{r}r'
\label{trans2}     \\
     -a^2 &=& g_{00}f'^2 - g_{rr}r'^2\ ,
\label{trans3}
\eea
where dot denotes partial differentiation with respect to $t$
(holding $\lambda$ fixed) while prime denotes partial differentiation
with respect to $\lambda$ (holding $t$ fixed).  

The object of transforming to the form Eq.~(\ref{line-cosmo}) is that the
bounding surface of the spherical mass distribution can be taken to be
at fixed $\lambda=\lambda_*$.  We wish now to make an identification
between the interior metric Eq.~(\ref{cosmo}) and the exterior metric
Eq.~(\ref{line-cosmo}) which is smooth, and such that this matching
surface is also a geodesic of the exterior (Schwarzschild-like) metric.
This is realized if the following conditions hold at the boundary:
\bea
     r(t,\lambda_*) &=&  \lambda_* a(t)
\label{match1}     \\
     r'(t,\lambda_*) &=& a(t)\ ,
\label{match2}     \\
     N(t,\lambda_*) &=&  1
\label{match3}     \\
     N'(t,\lambda_*) &=& 0\ .
\label{match4}
\eea
Using Eqs.~(\ref{match1}) and~(\ref{match2}) as the boundary condition,
one can integrate Eqs.~(\ref{trans2}) and~(\ref{trans3}) to arrive at
the complete coordinate transformation $T=f(t,\lambda)$ and
$r(t,\lambda)$, for arbitrary functions $g_{00}(r)$ and $g_{rr}(r)$.
For Eqs.~(\ref{match3}) and~(\ref{match4}) to be satisfied, conditions
need to be placed on $g_{00}(r)$ and $g_{rr}(r)$:
\bea
    \dot{r}^2(t,\lambda_*) &=& 1 - g_{rr}^{-1}(r(t,\lambda_*))  \\
    g_{00}g_{rr} &=& E^2 = {\rm constant}\ ,
\eea
for Eqs.~(\ref{match3}) and~(\ref{match4}) respectively.
One can quickly confirm that these expressions combined imply
$\lambda = \lambda_*$ follows a geodesic worldline and, moreover,
these expressions allow us to determine the metric components of
Eq.~(\ref{line-static}) uniquely from a given $a(t)$:
\bea
     g_{00} &=& E^2(1 - \lambda_*^2\dot{a}^2)
\label{metric-00}     \\
     g_{rr}^{-1} &=& 1 - \lambda_*^2\dot{a}^2\ ,
\label{metric-rr}
\eea
with $r(t,\lambda_*) = \lambda_*a(t)$.  

Thus, by requiring our new gravitational physics to respect Birkhoff's
law, the metric around a spherically symmetric matter source is
completely specified by the cosmology $a(t)$.  To determine the
dependence of the metric on the source mass, one need only select a
different $\lambda_*$.  The remaining parameter $E$ may be specified
by an arbitrary choice of time normalization.  Notice that for
a general scale factor evolution $a(t)$, superposition and linearity
of the metric must be sacrificed, even in the weak-field limit.\footnote{
  Note that DGP gravity does not satisfy the condition that
  $g_{00}g_{rr} = {\rm constant}$ and, therefore, cannot satisfy
  a dynamical version of Birkhoff's law.  But, while DGP gravity
  does not fall under the category of modified-gravity theories
  under consideration, it does share many of the same properties,
  such as the lack of superposition and linearity, as well as
  properties to be noted in the coming sections.}
Let us see what consequences this has for phenomenology.

\section{Governing Scales}

The exterior (Schwarzschild-like) metric is given by
Eqs.~(\ref{metric-00}) and~(\ref{metric-rr}).  For the gravitational
force law to approach Einstein's at short distances, we require
\be
	g_{00}(r) = g_{rr}^{-1}(r) = 1 - {r_g\over r}\ ,
\label{ordinary-Einstein}
\ee
when $r$ is small and $r_g=2GM$ is the usual definition of the
Schwarzschild radius.  On the other hand, cosmology at early times
(but still during matter dominated regime) must
obey the conventional Friedmann equation
\be
     H^2 \equiv {\dot{a}^2\over a^2} = {8\pi G\over 3}\rho\ ,
\label{ordinary-Friedmann}
\ee
where we use $\rho(t)$ to denote the matter density.
The relationship between mass and $\lambda_*$ is therefore
\be
     M(\lambda_*) \equiv {r_g\over 2 G} =  {4\pi\over 3}\rho(t) r^3 =
     {4\pi\over 3}\lambda_*^3\rho a^3\ ,
\label{Mass}
\ee
where $\rho(t)$ is the matter density, and $\rho a^3$ is a constant
with respect to $t$ for a dust-filled universe.

How small should $r$ be, or how large should $H$ be, for this
conventional behavior (Eqs.~(\ref{ordinary-Einstein}) and
(\ref{ordinary-Friedmann})) to be applicable?  Clearly, the
cosmological evolution must be conventional when $H \gg H_0$, where
$H_0$ is today's Hubble scale (approximately the scale at which
acceleration sets in) or in other words, rearranging
Eq.~(\ref{metric-rr}), when
\be 
	{1-g_{rr}^{-1}\over r^2} \gg H_0^2\ .
\ee
But this implies that, just as
in the DGP example,\footnote{
  In DGP gravity, $r_*$ is the distance at which scalar would-be
  radion modes become free to propagate, adding a Brans-Dicke type
  scalar to the existing gravitational interactions, where
  the strength of the Brans-Dicke coupling, $\omega$, depends
  sensitively on the background cosmology.}
\be
	r_* = \left(r_g\over H_0^2\right)^{1/3}\ ,
\ee
and not $r_c=H_0^{-1}$, is the distance smaller than which one expects
deviations from Einstein gravity to be small, but larger than which
{\em significant} deviations from Einstein must occur in the force law
in order to reproduce the desired cosmic history as a modification of
the Friedmann equation.  Indeed, we may write our modified Friedmann
equation in the very general form:
\be
     H^2 = H_0^2~g\left(x\right)\ ,
\label{mod-Fried}
\ee
with the dimensionless quantity $x$ defined as
\be
     x \equiv {8\pi G\rho\over3H_0^2}\ .
\label{xdef}
\ee
$g(x)$ is such that $g(x) \rightarrow x$ when $x\gg 1$, but $g(x)$
substantially deviates from $x$ otherwise, e.g. for the cosmological
constant case $g(x)=x+\Omega_\Lambda$, with $x = \Omega_m(1+z)^3$ from
Eq.~(\ref{xdef}) and $\Omega_m$ and $\Omega_\Lambda$ the energy
densities of matter and cosmological constant in terms of critical
density at redshift $z=0$.  The modified Schwarzschild metric may then
be read from Eqs.~(\ref{metric-00}) and~(\ref{metric-rr}):
\be
     E^{-2}g_{00} = g_{rr}^{-1} =
1-r^2H_0^2g\left(x\right)\ .
\label{mod-Schwarz}
\ee
where now $x = r_*^3/r^3$, using the definition of $r_*$ and
Eq.~(\ref{Mass}).

\section{Growth of Perturbations}

We next wish to study the influence of the modified gravitational
force law on the evolution of perturbations in the universe.  Define
the top-hat overdensity $\delta(t)$ of a spherical mass of dust with mass
$M$ and radius $r$ by
\be
	1+\delta = {M\over {4\pi\over 3}\bar{\rho}r^3}\ ,
\label{delta1}
\ee
where $\bar{\rho}(t)$ is the background matter density (dust
component only, i.e., not including the energy density of the dark
energy).  The conventional evolution of $\delta(t)$ is governed by
the following equation \cite{Peebles}
\be
	\ddot{\delta}+2\bar{H}\dot{\delta}
	-{4\over 3}{1\over 1+\delta}\dot{\delta}^2
	= 4\pi G\bar{\rho}(1+\delta)\delta\ .
\label{de}
\ee
The quantity $\bar{H}(t)$ corresponds to the background evolution.  We
wish to compare this relationship with that for the same scale factor
evolution (and correspondingly, the same $\bar{\rho}(t)$ evolution),
but with modified gravity.

By exploiting the Birkhoff's law constraint, one may compute the
evolution of overdensities by merely following the geodesics of
spherical masses, without regard to physics outside the spherical mass
itself.  Note that this is not possible unless spherically symmetric
configurations respect the metric Eqs.~(\ref{metric-00})
and~(\ref{metric-rr}). (DGP gravity in particular does not fall under
this category of Schwarzschild-like metrics.)  Using the geodesic
equation as expressed by differentiating Eq.~(\ref{metric-rr}) with
respect to $t$, we get
\be
     \ddot{r} = -{1\over 2}{d\over dr} g_{rr}^{-1}
     = rH_0^2\left[g(x)-{3\over 2}xg'(x)\right]\ .
\label{geodesic}
\ee
Note that this resembles Newton's second law, but is fully
relativistic.  Using some algebra and rewriting Eq.~(\ref{delta1}) as
follows
\be
     1+\delta = x{3H_0^2\over 8\pi G\bar{\rho}}\ ,
\ee
where again $x = {r_*^3\over r^3}$, one may use Eq.~(\ref{geodesic})
to derive a new governing equation for $\delta(t)$
\bea
     \ddot{\delta}+2\bar{H}\dot{\delta}
     -{4\over 3}{1\over 1+\delta}\dot{\delta}^2
     = 3(1+\delta)H_0^2\left[
       {3\over 2}\bar{x}(1+\delta)g'(\bar{x}(1+\delta))
       - g(\bar{x}(1+\delta))\right] \nonumber \\
 -  3(1+\delta)H_0^2\left[{3\over 2}\bar{x}g'(\bar{x}) - g(\bar{x})\right]\ ,
\label{mg}
\eea
where we define $\bar{x} \equiv {8\pi G\bar{\rho}\over3H_0^2}=\Omega_m
\bar{a}^{-3}$.  The quantity $\bar{a}(t)$ is the background scale
factor.  Note that Eq.~(\ref{mg}) is dependent only on the
background evolution and makes no reference to the mass $M$ or the
radius $r_*$.  Evolution begins with $\bar{x} \gg 1$ and $\bar{x}$
decreasing with time.  Deviations from the usual Einstein evolution
occur when $\bar{x} \sim 1$.  

Equation~(\ref{mg}) is equivalent to Eq.~(50)
in~\cite{Multamaki:2003vs}, where it is derived assuming the
continuity equation and the Friedman equation {\em for
fluctuations}. It turns out that this is equivalent to assuming the
validity of Birkhoff's theorem. As we mentioned above, however, this
approach is not justified when dealing with DGP gravity; basically,
one cannot infer the evolution of a spherical perturbation from the
evolution of the scale factor.

In linear perturbation theory, we may simplify
Eq.~(\ref{mg}):
\be
     \ddot{\delta}+2\bar{H}\dot{\delta}
     = 4\pi G\bar{\rho}\delta\left[
       g'(\bar{x}) + 3\bar{x}g''(\bar{x})\right]\ .
       \label{LPT}
\ee
If $g(x) = x +A_1 + A_2~x^{2/3}$, where $A_1$ and $A_2$ are constants, 
we recover the usual scenario
Eq.~(\ref{de}) and modified gravity yields an identical answer to the
usual dark energy scenario.  Choosing the time variable to be
$\bar{x}$, we see that
\be
   3\bar{x}g(\bar{x}){d^2\delta\over d\bar{x}^2}
   + \left[g(\bar{x})+{3\over 2}\bar{x}g'(\bar{x})\right]
   {d\delta\over d\bar{x}} = {1\over 2}
   \left[g'(\bar{x}) + 3\bar{x}g''(\bar{x})\right]\delta\ .
\label{dlinMG}
\ee
Compare this expression with the corresponding one for evolution in a
dark energy background from Eq.~(\ref{de})
\be
   3\bar{x}g(\bar{x}){d^2\delta\over d\bar{x}^2}
   + \left[g(\bar{x})+{3\over 2}\bar{x}g'(\bar{x})\right]
   {d\delta\over d\bar{x}} = {1\over 2}\delta\ .\ \ \ \ \ {\rm (dark~energy)}
\label{dlinDE}
\ee
It is interesting to note that, unlike Eq.~(\ref{dlinDE}), the case of
modified gravity Eq.~(\ref{dlinMG}) has a decaying mode solution
$D_{-}^{\rm MG} \propto \bar{H} \propto g^{1/2}$ for {\em arbitrary}
expansion histories. As a result of this, one finds that the growing
mode $D_+^{\rm MG}$ obeys
\be
D_+^{\rm MG}(\bar{x}) = \frac{5}{6}\Omega_m^{1/3}g^{1/2}\
\int_{\bar{x}}^\infty \frac{dx'}{[g(x')]^{3/2} x'^{1/3}}
\label{Dplus}
\ee
where $D_+^{\rm MG}$ is normalized so that it scales like the scale
factor at early times, $\bar{x}\gg1$.  When the expansion history can
be described by a linear combination of dust, vacuum energy and
curvature, $g(x) = x +A_1 + A_2~x^{2/3}$, dark energy and modified
gravity give rise to the same linear growth of density perturbations
and Eq.~(\ref{Dplus}) becomes the standard quadrature representation
\cite{Heath}. For a general expansion history, one cannot use
Eq.~(\ref{Dplus}) to find the growing mode in a dark energy background
$D_+^{\rm DE}$ and must solve Eq.~(\ref{dlinDE}) instead.

\begin{figure}[h!]
\centerline{\epsfxsize=9cm\epsffile{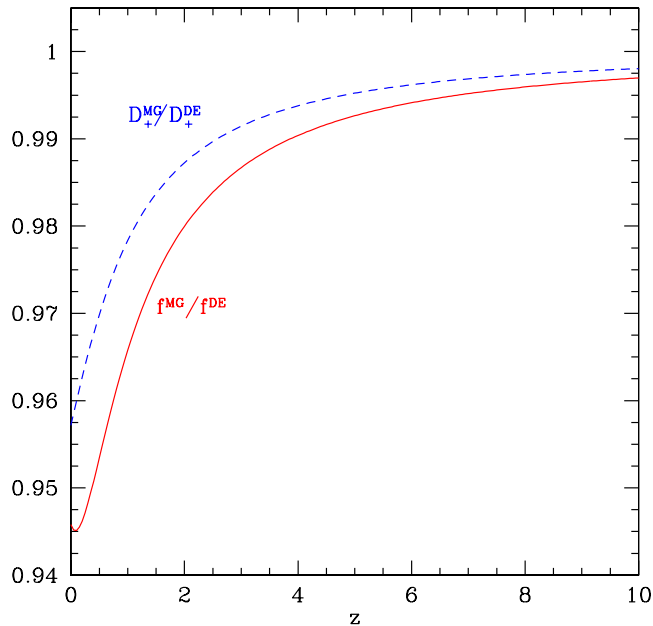}}
\caption{
Growth of perturbations as a function of redshift in dark
energy (DE) models and modified gravity (MG) models with {\em
identical expansion histories}, given by Eq.~(\ref{gGT}). The dashed
line shows the ratio of density perturbation growth factors $D_+^{\rm
MG}/D_+^{\rm DE}$, the solid line shows the ratio of the velocity
perturbation growth factors $f^{\rm MG}/f^{\rm DE}$.
}
\label{growth}
\end{figure}

Let us now consider how the linear growth of perturbations can
distinguish between dark energy and modified gravity models with the
{\em same} expansion history. This is relevant because in the near
future, planned experiments will measure the expansion history from
$z=0$ to $z \simeq 2$ with very good accuracy (see
e.g. \cite{Frieman,Karim}).  For other tests of gravity using
large-scale structure see e.g.~\cite{Francis,Enrique}.  To illustrate
our results, we assume a modified Friedman equation given by

\be
g(x) = \Big[ c + \sqrt{c^2 + x} \Big]^2,
\label{gGT}
\ee
where $c$ is a constant, $a=(1-\Omega_m)/2$. This corresponds to
``extra-dimensions-inspired" modified gravity in~\cite{Dvali:2003rk}
when their $\alpha=1$ (which corresponds to an effective equation of
state with $w_{\rm eff} \simeq -0.7$ for $z$ between 0 and 2). In
all calculations we set $\Omega_m=1/3$ for simplicity.

Figure~\ref{growth} shows the results for the ratio $D_+^{\rm
MG}/D_+^{\rm DE}$ (dashed line) as a function of redshift $z$. The
modified gravity perturbations growth slower, and the decline of this
ratio as redshift approaches zero is a general feature, not restricted
to Eq.~(\ref{gGT}), that can be understood in general terms. From
Eqs.~(\ref{LPT}-\ref{dlinDE}) we see that for the same expansion
history, the difference in growth can be thought as coming from an
effective gravitational constant $G[g'(\bar{x}) +
3\bar{x}g''(\bar{x})]$. At high redshift $g(\bar{x}) \approx \bar{x}$,
but at low redshift the constraint that the universe accelerates
($\ddot{\bar{a}}>0$) implies ${\rm d}\ln g/{\rm d}\ln \bar{x} < 2/3$. If we
model $g(x)$ as a local power-law, $g(\bar{x}) \sim \bar{x}^n$, where
$n$ changes very slowly from $n=1$ at high redshift to $n<2/3$ when $z
\sim 1$, we see that $g'(\bar{x}) + 3\bar{x}g''(\bar{x}) \sim
n(3n-2)g(\bar{x})/\bar{x}$, so as $n$ becomes closer to $2/3$,
$g'(\bar{x}) + 3\bar{x}g''(\bar{x})$ becomes small. Although in
practice one must take into account the time dependence of $n$, this
illustrates why the growth of structure is slower in the case of
modified gravity.  As long as we are only concerned with linear
perturbations around a cosmological background, this effective Newton's
constant is wavelength independent.  However, when perturbations
become strong, and self-gravitation dominates cosmology, we will see
that the effective $G$ does acquire distance dependence (see Sec.~VI).

Figure~\ref{growth} also shows the ratio $f^{\rm MG}/f^{\rm DE}$,
where $f\equiv {\rm d}\ln D_+ /{\rm d}\ln \bar{a}$ governs the growth of
velocity fluctuations in linear perturbation theory, that is, the
velocity divergence evolves as $\theta = -\bar{H}\bar{a}f \delta$ from the
linearized continuity equation (see e.g., Ref.~\cite{Peebles}). We see
that slower growth leads to smaller time derivative, thus $f^{\rm
MG}/f^{\rm DE}$ also decreases as $z \rightarrow 0$. These deviations
are detectable with precision measures of large-scale structure; for
example, $D_+$ and $f$ can be derived from joint measurements of the
redshift-space power spectrum anisotropy and bispectrum. The Sloan
Digital Sky Survey should be able to probe these quantities with
statistical errors of order a few percent \cite{Colombi1998}.

So far we have discussed the linear growth of
perturbations. Equation~(\ref{mg}) can be recast in a more useful way
to study the non-linear evolution in perturbation theory,
\bea
   3\bar{x}g(\bar{x}){d^2\delta\over d\bar{x}^2}
   + \left[g(\bar{x})+{3\over 2}\bar{x}g'(\bar{x})\right]
   {d\delta\over d\bar{x}} -4 \frac{\bar{x}g(\bar{x})}{1+\delta}
\Big(\frac{d\delta}{d\bar{x}}\Big)^2 
  & = & \Big( \frac{1+\delta}{2x} \Big)\ \sum_{n=1}^\infty
\frac{(\delta \bar{x})^n}{n!}  \nonumber \\
   & & \times  \Big[(3n-2) g^{(n)}+3\bar{x} g^{(n+1)} \Big]
\label{MGnl}
\eea
A couple of points are worth stressing here. First, this equation
becomes identical to that in the standard dark energy scenario,
Eq.~(\ref{de}), only if $g(x)=A_1+x$, i.e. a curvature term is no
longer degenerate as in the linear case. Expanded to second order,
this equation can be used to compute the skewness of the density field
(see also~\cite{Multamaki:2003vs}); we have checked that for the same
expansion history given by Eq.~(\ref{gGT}) the skewness changes by
less than one percent in modified gravity compared to dark energy
models. Only a rather large third derivative of $g$ can induce an
appreciable change in skewness from that in the dark energy case.
Such extreme models have been studied in~\cite{Multamaki:2003vs}
(e.g. using Eq.~(\ref{mpc}) below with $n=0$ and $q=5$); we have
verified that in this case the skewness can change by up to
$10\%$ compared to dark energy models, and also $D_+^{\rm MG}/D_+^{\rm
DE}$ can be {\em larger} than unity (up to $1.2$ at $z=0$) due to a
fast variation of $g$ with $\bar{x}$.

Ultimately, unless the source of acceleration is a vacuum energy
component, determination of the expansion history of the universe plus
the growth of structure should allow us to identify, through the
divergence between Eq.~(\ref{de}) and Eq.~(\ref{mg}), whether today's
cosmic acceleration may be attributed to modified gravity or dark
energy.

\section{Evolution of Gravitational Potentials: The ISW Effect}

The microwave background provides another window to differentiate 
between dark energy and modified gravity through the late-time integrated 
Sachs-Wolfe (ISW) effect  due to the decay in gravitational potentials 
at late times~(see e.g. \cite{HuDodelson}). In order to assess how a 
modified gravitational force law affects
the ISW effect, we must find the evolution of the cosmological
gravitational potentials, $\Psi(t,\lambda)$ and $\Phi(t,\lambda)$,
which are defined using the line element
\be
     ds^2 = \left[1+2\Psi(t,\lambda)\right]dt^2
     - \bar{a}^2(t)\left[1+2\Phi(t,\lambda)\right]
     \left[d\lambda^2+\lambda^2d\Omega\right]\ ,
\label{potentials}
\ee
where $\bar{a}(t)$ is the background scale factor evolution.  We
are only interested in potentials and overdensities that are small
and, therefore, only interested in linear perturbations around the
cosmological background.

We have the equations for the evolution of a linear top-hat
overdensity $\delta(t)$ of radial extent $\lambda_*$,
Eq.~(\ref{dlinMG}), and we know the complete metric for
that matter configuration:
\be
     ds^2 = dt^2 - \bar{a}^2(t)\left[{d\lambda^2\over 1-\kappa\lambda^2}
	 + \lambda^2d\Omega\right]\ ,
\label{curvature}
\ee
where $a^3 = (1+\delta)\bar{a}^3$, and $\kappa$ is the curvature parameter
associated with the magnitude of the overdensity.  Using $r = a\lambda_*$
and integrating Eq.~(\ref{geodesic}), we arrive at
\be
     \dot{r}^2 = (1-\kappa\lambda_*^2) - g_{rr}^{-1}\ ,
\ee
where the constant of integration is determined by smoothly
connecting a background cosmological expansion outside $\lambda_*$ with
an overdense space inside $\lambda_*$.  Then, one may deduce that
\be
     \kappa = \bar{a}^2H_0^2\bar{x}g'(\bar{x})\delta
          + {2\over 3}\bar{a}\dot{\bar{a}}\dot{\delta}\ .
\ee
One can confirm using Eq.~(\ref{Dplus}) that $\kappa$ is indeed a
constant.

One needs only identify a coordinate transformation taking the metric
Eq.~(\ref{curvature}) into the form Eq.~(\ref{potentials}) and
read off $\Psi$ and $\Phi$.  After some algebra we find
\bea
     \Psi(t,\lambda) &=& -{\bar{a}^2\lambda_*^2\over 6}
     \left(1-{\lambda^2\over\lambda_*^2}\right)
     \left[4\pi G\bar{\rho}\delta(g'+3\bar{x}g'')\right]  \\
     \Phi(t,\lambda) &=& {\bar{a}^2\lambda_*^2\over 6}
     \left(1-{\lambda^2\over\lambda_*^2}\right)
     \left[4\pi G\bar{\rho}\delta g'\right]\ , 
\eea
when $\lambda < \lambda_*$ and the potentials vanish outside $\lambda_*$.
By superposing top-hat linear overdensities, we surmise that for
general $r$--dependent $\delta$
\bea
     \nabla^2\Psi &=& 4\pi G\bar{\rho} \bar{a}^2
     \left[g'(\bar{x})+3\bar{x}g''(\bar{x})\right]\delta    \label{PoissonPhi} \\
     \nabla^2\Phi &=& -4\pi G\bar{\rho}\bar{a}^2g'(\bar{x})\delta\ , \label{PoissonPsi}
\eea
where the Laplacian is with respect to comoving coordinates.  These
expressions for the linear gravitational potentials are the generalization of the usual result for Einstein gravity, around a specific cosmological background.  One arrives at time-dependent effective Newton's constants (in general, different for $\Psi$ versus $\Phi$).  At early times, $G_{\rm eff} \rightarrow G$ but deviates from the true Newton's constant significantly as $\bar{H}\rightarrow H_0$.  Note that this large discrepancy only applies to self-gravitation of linear density perturbations.  For example, solar system manifestations of Newton's constant, in this context, are {\em not} linear and will only have small deviations from the true Newton's constant, as will be seen in the next section.

We may now take Eqs.~(\ref{PoissonPsi}--\ref{PoissonPhi}) and apply them to ascertain the ISW effect on the cosmic microwave background, which is proportional to the integral of $\dot{\Phi}-\dot{\Psi}$ along the line of sight ~(see e.g. \cite{HuDodelson}). From Eqs.~(\ref{PoissonPhi}--\ref{PoissonPsi}) we find, after a Fourier transformation,
\bea
A(\dot{\Phi}-\dot{\Psi})^{\rm DE}&=& (1-f^{\rm DE})\  D_+^{\rm DE} \label{potDE} \\
A (\dot{\Phi}-\dot{\Psi})^{\rm MG}&=& \Big[(1-f^{\rm MG})(g'+\frac{3}{2}\bar{x}g'')+\frac{3}{2}(5\bar{x}g''+3\bar{x}^2g''')\Big]
\  D_+^{\rm MG} \label{potMG}\ ,
 \eea
where $A\equiv -k^2 /(8\pi G \bar{H}\bar{\rho}\bar{a}^2 \delta_0)$ with $k$ the comoving wavenumber and $\delta_0$ the amplitude of density perturbations at some early time. The second term in Eq.~(\ref{potMG}), representing the time derivative of the effective gravitational constant, leads to an additional decay of the potentials that can be quite significant at low redshifts. 

\begin{figure}[t!]
\centerline{\epsfxsize=9cm\epsffile{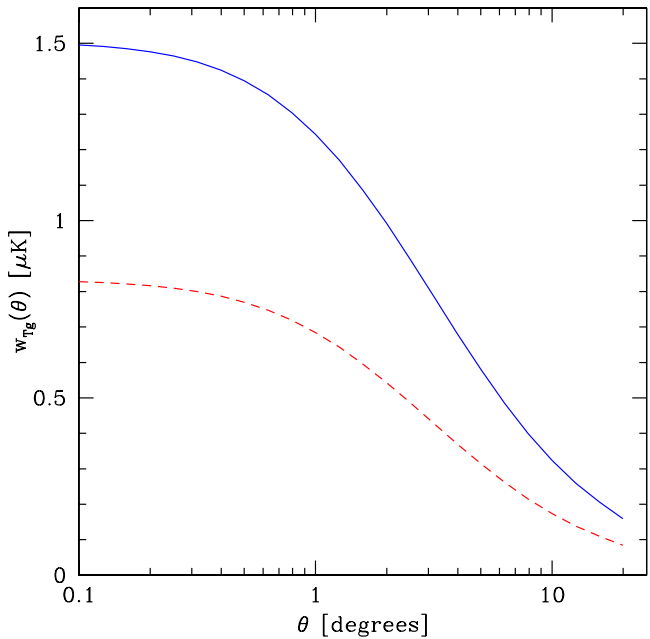}}
\caption{The angular cross-correlation function between galaxies and CMB temperature anisotropies. The solid line shows the dark energy case, dashed line corresponds to modified gravity with the same expansion history.}
\label{SDSSWMAP}
\end{figure}

We compute the cross-correlation between temperature anisotropies due to the ISW effect and galaxy fluctuations, which has recently been detected using the WMAP and SDSS datasets~\cite{FGC,Scranton}. The angular cross-correlation function $w_{Tg}(\theta)$ can be written as
\be
w^{\rm DE}_{Tg}(\theta) = 3T_0\Omega_m b (2\pi)^2 \frac{H_0^3}{c^3}\int dz \sqrt{g}\ D_+^2 (1-f)\ w_g(z) \int \frac{dk}{k} P(k) J_0(k\theta\chi)\ ,
\label{wde}
\ee
for the dark energy case, whereas  for modified gravity 
\bea
w^{\rm MG}_{Tg}(\theta) &=& 3T_0\Omega_m b (2\pi)^2 \frac{H_0^3}{c^3}\int dz \sqrt{g}\ D_+^2 \Big[(1-f) (g'+\frac{3}{2}\bar{x}g'')+\frac{3}{2}\bar{x}(5g''+3\bar{x}g''')\Big]
\ w_g(z) \nonumber \\ & & \times \int \frac{dk}{k} P(k) J_0(k\theta\chi) \ ,
\label{wmg}
\eea
where we have assumed the small-angle approximation, and $T_0$ is the CMB temperature, $b$ is the linear bias of LRG galaxies, $w_g(z)$ is the galaxy selection function, $P(k)$ is the dark matter power spectrum at $z=0$, $J_0$ is a Bessel function, and $\chi(z)$ is the comoving distance as a function of redshift $z$.   We assume, following the results in~\cite{Scranton} (see their Fig.~3), that\footnote{This choice is a matter of convention here since it scales both predictions by the same amount. In practice, by measuring the angular bispectrum of LRG galaxies one can determine $b$, which due to the results in the previous section should be almost independent of whether DE or MG is present.} $b=5.47$, and use their selection function for the $z=0.49$ sample.

Figure~\ref{SDSSWMAP} shows the results, again using the expansion history given by Eq.~(\ref{gGT}). We see that the MG angular correlation function is suppressed by a factor of about two compared to the case of DE, for the same expansion history.  This anomaly is a reflection of the order-unity anomalous effective Newton's constant seen in Eqs.~(\ref{PoissonPsi}--\ref{PoissonPhi}) when the Hubble parameter is
near today's value.  Although the ISW anomaly is not yet detectable, it should be so in the near future with the completion of the SDSS survey or other future probes of structure at intermediate redshifts.

The suppression of the late-time ISW effect that results from MG can help explain, at least partially, the low amplitude of the CMB power spectrum at low multipoles as confirmed by WMAP~\cite{Spergel}.  It is worth emphasizing that these results depend on the specific prescription of exploiting Birkhoff's law for understanding modified gravity from cosmological physics, and it will be important to check whether the large ISW discrepancy carries over to more general prescriptions of modified gravity.

\section{Orbit Precession}

We have seen that modification of the Friedmann equation leads to a
modification of the gravitational force law.  This modification of the
force law may, in turn, lead to small-but-detectable corrections at
distances much smaller than today's Hubble scale.  Indeed, under some
circumstances (e.g., if no physics other than that deduced from
cosmological evolution emerges at small scales) these alterations of
cosmology on the largest observable distances may cause observable
deviations from known physics at even solar system scales.  Let us
elaborate.

The precession $\Delta\phi$ of the perihelion per orbit in the
background of a metric of the form Eq.~(\ref{line-static}) may be
determined in the usual way:
\be
     \Delta\phi = \int dr {J\over r^2}\sqrt{g_{00}g_{rr}
       \over E^2-g_{00}(1+J/r^2)}\ ,
\ee
where $E = N^2\dot{T}$, $J = r^2\dot{\phi}$, and dot refers to
differentiation with respect to proper time.  For a nearly circular
orbit with a metric of the form Eq.~(\ref{mod-Schwarz}), one may
compute the precession rate:
\be
     {d\over dT}\Delta\phi = {H_0\over\sqrt{2}}x^{1/2}
     \left[\sqrt{g-{3\over 2}xg'\over 4g-{9\over 2}xg' + {9\over 2}x^2g''}
       -1\right]\ ,
\label{precession}
\ee
with again, $x = {r_*^3\over r^3}$.  The leading contribution to orbit
precession from the altered metric Eq.~(\ref{mod-Schwarz}) comes from
the simple alteration of the Newtonian potential.

We are particularly interested in orbits whose radii around a central
body of Schwarzschild radius $r_g$ are much smaller than $r_* =
(r_g/H_0^2)^{1/3}$.  Then, corrections from the modified gravity are
small.  One may represent the function $g(x)$ as
\be
     g(x) = x\left[1+\delta g(x)\right], 
\ee
with $\delta g(x) \ll {\cal O}(1)$.  Then, Eq.~(\ref{precession})
reduces to
\be
     {d\over dT}\Delta\phi = {6H_0\over\sqrt{2}}x^{3/2}
     \left(\delta g'+{3x\over 4}\delta g''\right)\ ,
\label{precession-small}
\ee
to leading order in $\delta g$.  Recall that here $x \gg 1$.

Take as an instructive example the form of the modified Friedmann
equation found in Cardassian models \cite{Freese:2002gv}.  This can be
written as
\be
	g(x) = x\left[1+cx^{-q(1-n)}\right]^{1/q}\ ,
\label{mpc}
\ee
where $n$ and $q$ are parameters of the modification and where
\be
     c = \left({(1+z_{\rm eq})^{3q}
       \over 1+(1+z_{\rm eq})^{3q(1-n)}}\right)^{1-n}\ .
\ee
The quantity $z_{\rm eq}$ is the redshift at which the two terms
inside bracket in Eq.~(\ref{mpc}) are equal, a quantity of order
unity.  Then, using Eq.~(\ref{precession-small}), the anomalous orbit
precession rate is
\be
     {d\over dT}\Delta\phi = {3cH_0\over 2\sqrt{2}}(n-1)[3q(n-1)+1]
          \left({r^3\over r_*^3}\right)^{q(1-n)-{1\over 2}}\ .
\label{precession-mpc}
\ee
When $q(1-n)={1\over 2}$, this expression is independent of the radius
of the orbit, or the mass of the central body, and the anomalous
precession rate is proportional to today's Hubble parameter, $H_0 \sim
10~\mu{\rm as/year}$.  Such a precession rate is on the threshold of
detection by precision ephemeris measurements of the inner solar
system, particularly with intriguing developments this decade coming
from two Mercury-bound missions (BepiColombo and MESSENGER) as well as
improvements in lunar ranging observations
\cite{Lue:2002sw,Dvali:2002vf,Nordtvedt:ts,Williams:1995nq,Milani:2002hw,Will:2001mx}.
When $q(1-n) \ne {1\over 2}$, there is a relevant distance-dependence
for the anomalous precession rate, where the governing length scale is
once again $r_*$.  Since we are primarily concerned with orbits such
that $r\ll r_*$, the dimensionless distance factor in the anomalous
precession rate, $(r^3/r_*^3)^{q(1-n)-{1\over 2}}$, will either be
huge or tiny.  Thus, with solar system constraints in mind, the
parametric range where $q(1-n) \lesssim {1\over 2}$ can already be
ruled out.  However, for $q(1-n) \gg {1\over 2}$, no solar system test
are likely to discover discrepancies based on anomalous orbit
precession in the foreseeable future.

\section{Concluding Remarks}

In this paper we showed how modifying gravity to effect the
observed late-time cosmological acceleration at scales of today's Hubble
radius, $H_0^{-1}$, can lead naturally to corresponding modifications of
gravitational interactions at scales much shorter than $H_0^{-1}$.
Indeed, by presuming that the new gravitational physics obeys
a limited version of Birkhoff's law, we were able to derive the
precise form of the modifications to Newton's law of gravitation at
short (sub-cosmological) distances.  We then showed that an observer
in the gravitational field of a central source whose Schwarzschild
radius is $r_g$, experiences substantial deviations from the usual
Schwarzschild metric at all distances greater than approximately

\be
     r_* = \left({r_g\over H_0^2}\right)^{1/3}\ .
\ee
For many models these deviations will be measurable through
observation of orbital precession of solar system objects in the
coming decade.  We also discussed the evolution of density
perturbations and showed that, unless the acceleration of the universe
is driven by an effective vacuum energy, simultaneous measurement of
the expansion history and growth of large-scale structure can be used
to distinguish modified gravity from dark energy.  In addition, the
cross-correlation of galaxy distribution and the cosmic microwave
background temperature anisotropy can detect anomalies in the
late-time integrated Sachs--Wolfe effect caused by modified gravity.
Such measurements will be available imminently.

It is instructive that these results are identical to those found for
the braneworld theory of Dvali, Gabadadze, and Porrati (DGP), even
though DGP gravity does not respect any dynamical version of
Birkhoff's law.  The correspondence between the scales of departure
from Einstein gravity in DGP and the Birkhoff's Law theories seems not
to be a coincidence.  Indeed, one suspects that it is quite general.
Imagine cosmology at extremely late times, when all matter surrounding
a particular gravitational source is swept away.  Then, if one
believes that an isolated, central source has a (quasi)static metric
description, it should be Schwarzschild at short distances and deviate
from Schwarzschild at large-enough distances.  How large?  Since this
metric must still encode the cosmology within it, i.e., test observers
at large distances from the source should recede from the source in
the manner dictated by the given late-time (accelerating) cosmology,
the {\em empty space} metric must include a repulsive force at
distances where cosmological flow overcomes the local gravitational
binding of the central source.  Thus, we expect substantial deviations
from Schwarzschild (in the form of a repulsive force) at distances
${\cal O}(r_*)$, Thus, it is not unreasonable to expect general
modifications of cosmology at today's Hubble scale to inevitably
affect local gravitational interactions at distances governed by
$r_*$; it is only the precise functional form of the deviations that
will vary from model to model.

\acknowledgements
We wish to thank K.~Benabed and T.~Vachaspati for helpful discussions,
and E.~Gazta\~naga and M.~Zaldarriaga for encouraging us to look at the
ISW effect. We also thank R.~Scranton for providing us with the selection
functions for the SDSS LRG samples.  A.~L. is grateful for the hospitality of
the Center for Cosmology and Particle Physics (New York University).  This
work is sponsored by DOE Grant DEFG0295ER40898, the CWRU Office of
the Provost, NASA grant NAG5-12100, and NSF grant PHY-0101738.

\end{document}